\documentclass[12pt, a4paper]{article}

\usepackage[T1]{fontenc}
\usepackage[utf8]{inputenc}
\usepackage[margin=2.5cm]{geometry}
\usepackage{setspace}
\usepackage{xcolor}
\usepackage{enumitem}
\usepackage{hyperref}
\usepackage{parskip}
\usepackage{fancyhdr}
\usepackage{mdframed}
\usepackage{listings}
\usepackage{authblk}
\usepackage{booktabs}
\usepackage{array}
\usepackage{natbib}

\definecolor{slate}{RGB}{88, 103, 120}
\definecolor{slatelight}{RGB}{237, 240, 244}
\definecolor{codebg}{RGB}{248, 248, 248}
\definecolor{darkgrey}{RGB}{74, 74, 74}

\hypersetup{
  colorlinks=true,
  linkcolor=slate,
  citecolor=slate,
  urlcolor=slate,
  pdftitle={AI as Consumer and Participant},
  pdfauthor={Siyuan Ji}
}

\lstdefinestyle{sysml}{
  backgroundcolor=\color{codebg},
  basicstyle=\ttfamily\small,
  breaklines=true,
  frame=single,
  framerule=0.4pt,
  rulecolor=\color{darkgrey},
  xleftmargin=8pt, xrightmargin=8pt,
  aboveskip=8pt, belowskip=8pt
}

\newmdenv[
  backgroundcolor=slatelight,
  linecolor=slate, linewidth=1.2pt,
  innerleftmargin=14pt, innerrightmargin=14pt,
  innertopmargin=12pt, innerbottommargin=12pt,
  skipabove=16pt, skipbelow=16pt
]{thesisbox}

\newmdenv[
  backgroundcolor=white,
  linecolor=darkgrey, linewidth=0.5pt,
  leftline=true, rightline=false,
  topline=false, bottomline=false,
  innerleftmargin=10pt, innerrightmargin=8pt,
  innertopmargin=6pt, innerbottommargin=6pt,
  skipabove=10pt, skipbelow=10pt
]{exchange}

\newmdenv[
  backgroundcolor=slatelight,
  linecolor=slate, linewidth=0.8pt,
  innerleftmargin=14pt, innerrightmargin=14pt,
  innertopmargin=12pt, innerbottommargin=12pt,
  skipabove=14pt, skipbelow=14pt
]{contrastbox}

\setlist[itemize]{leftmargin=1.5em,itemsep=0.3em,topsep=0.4em}
\setlist[enumerate]{leftmargin=1.5em,itemsep=0.3em,topsep=0.4em}

\begin{document}

\title{%
  \textbf{AI as Consumer and Participant:}\\[0.3em]
  \textbf{A Co-Design Agenda for MBSE Substrates and Methodology}
}

\author{Siyuan Ji}
\affil{Wolfson School of Mechanical, Electrical and Manufacturing Engineering,
  Loughborough University, Loughborough, UK\\
  \texttt{s.ji@lboro.ac.uk}}

\date{}
\maketitle

\begin{abstract}
AI tools are being deployed over MBSE models today, and those models were not
designed for this kind of consumption. The problem is not simply that tools
hallucinate: well-prompted frontier models produce competent, useful output over
a conformant SysML model, but the reasoning they produce is drawn from training
rather than retrieved from the model itself, and different tools over the same
model produce different results with nothing in the record to adjudicate between
them. The model, in other words, is functioning as a prompt rather than as a
knowledge base. Attaching better tools to the same model does not resolve this.
The model and the methodology that governs its construction need to be designed
together for AI participation, treating the model as a machine-queryable knowledge
substrate rather than a structured artefact for human navigation, and that
co-design has not yet happened in any systematic way. This paper works through a
concrete workflow scenario to show what that gap looks like in practice, proposes
three principles that jointly characterise what model and methodology must achieve
together, and closes with a call to the community to begin this work before the
architectural decisions about AI integration settle without the methodological
foundation they require.
\end{abstract}


\bigskip

\section{Introduction}
\label{sec:intro}

For two decades the MBSE community has worked to replace the document-based
engineering record with model-based artefacts that every stakeholder can access,
query, and trust. The promise has been a single source of truth: a coherent
representation of a system supporting communication across disciplines,
traceability across lifecycle phases, and rigour in design decisions. Adoption
has grown, but the promise remains substantially unrealised. In most organisations
the MBSE model is a tool of the systems engineering minority, read by those who
speak SysML and largely opaque to the broader engineering stakeholders whose
decisions it was meant to inform~\citep{henderson2024adoption,chami2018mbse,campo2023mbse}.

A new kind of participant is now arriving. Large Language Models are entering
engineering workflows with the capacity to read SysML, answer natural language
questions about models, support agentic reasoning, and contribute to model
construction itself. They are consumers of the engineering record in ways that
earlier tools were not, drawing on that record to construct reasoning, answer
questions, and generate new content. And at higher autonomy levels, as the
Co-Pilot research roadmap makes clear~\citep{zhang2026copilot}, they are
participants in the modelling process itself, engaged continuously in workflows
rather than responding to isolated queries. The community's response has been
to intervene at the tool vertex of the MBSE triangle: attaching LLM capabilities
to existing modelling environments while language and methodology remain largely
as they were. This paper argues that is the wrong place to stop.

\begin{thesisbox}
\noindent\textbf{\large Position}\\[0.5em]
\textit{AI operates in MBSE both as a consumer of the knowledge substrate and
as a participant in the modelling workflow. Both roles impose requirements that
current MBSE practice does not meet, and the two sets of requirements are
co-dependent: the substrate cannot be designed for reliable AI consumption
without knowing what the methodology requires of it at each workflow stage, and
the methodology cannot govern AI participation without specifying what the
substrate must carry. Addressing one without the other produces a practice that
is either capable but ungoverned, or governed but unreliable. The community must
pursue co-design of substrate and methodology together. Neither is currently
being pursued. This paper articulates what co-design requires and calls the
community to act.}
\end{thesisbox}

\noindent
The argument proceeds as follows. Section~\ref{sec:triangle} establishes why
the current pattern of AI integration, concentrated at the tool vertex, leaves
the two most consequential vertices unaddressed and what structurally follows.
Section~\ref{sec:failure} grounds the co-design gap through a concrete workflow
scenario and a representative artefact drawn from the SysML~v2 specification
itself. Section~\ref{sec:principles} articulates three co-design principles
that jointly constrain substrate and methodology. Section~\ref{sec:codesign}
identifies who must lead this work and what must change in practice.
Section~\ref{sec:safety} addresses safety-critical concerns directly.
Section~\ref{sec:action} closes with a three-part call to act directed at
INCOSE, OMG, and the research community.

\section{The Tool-Vertex Trap}
\label{sec:triangle}

The MBSE triangle is a claim about co-dependence, not merely a classification.
A language without appropriate tooling is unusable in practice. Tooling without
methodology produces artefacts whose quality depends on individual judgement rather
than governed process. Methodology without a language capable of expressing what
the process requires produces compliant-but-empty artefacts. Each vertex constrains
and enables the others; a change at one vertex that leaves the others unchanged
produces an incoherent practice.

The AI-for-MBSE literature is, almost without exception, a tool-vertex intervention.
Retrieval-augmented generation systems expose existing models to
LLMs~\citep{gadewadikar2026rag}. Plug-in architectures embed LLM capabilities
directly into commercial modelling environments, sequencing tasks with
human-in-the-loop review~\citep{zhang2026copilot}. Agentic frameworks model
engineering workflows as constrained sequential decision processes over existing
toolchains~\citep{son2026aei}. Maturity models chart levels of AI integration from
tool-supported to autonomous~\citep{bernijazov2025aimb}. These are genuine
contributions that establish what AI can do at the tool level. The question none
of them addresses is what the other two vertices must provide for that capability
to be reliable.

The Co-Pilot roadmap is the most explicit treatment of higher-autonomy AI
engagement in MBSE, and its treatment of workflow is illuminating precisely
because of where it stops. The roadmap acknowledges that its core capability
groups are expected to compose into workflows at higher autonomy levels, and
its case study demonstrates a tool plug-in that sequences modelling tasks with
human-in-the-loop governance~\citep{zhang2026copilot}. This is workflow in the
operational sense: a sequence of tool-mediated steps. What is absent is the
methodological sense of workflow: the specification of what the artefact must
contain at the point of AI engagement, under what conditions an AI contribution
is valid, and how AI-authored content is distinguished from human-authored content
in the engineering record. These are process definition questions of exactly the
kind a modelling methodology is supposed to answer. The roadmap correctly identifies
them as future work at higher autonomy levels. The point of this paper is that
they cannot be deferred: they are structurally prior to reliable AI participation,
and they require co-design of the substrate with the methodology.

The INCOSE SE Vision 2035 frames the same gap at the level of aspiration. It
envisions automated and efficient workflows, AI-enabled toolchains, and
data-driven context-aware algorithms~\citep{incosevision2035}. AI is treated
throughout as a tool capability to be integrated into existing practice, not
as a participant whose involvement the practice must be designed around. The
methodology vertex is conspicuously absent from the Vision's treatment of AI.
The MBSE 2.0 framing identifies language, tools, and methodology as vertices all
requiring evolution~\citep{mbse2025twopoint}, but its AI integration proposals
address intelligent co-design as an LLM-enabled tool capability rather than as
a question about what the methodology must specify when AI becomes a participant
in the modelling process.

The structural consequence of tool-vertex-only intervention is this. When AI
operates within a workflow at higher autonomy levels, the substrate it engages
with at each workflow stage determines what it can reliably reason over or
contribute. If the methodology specifies nothing about substrate state at the
moment of AI engagement, the AI's behaviour at that stage is ungoverned: it will
draw on training to fill what the substrate does not contain, different instances
will produce different results, and the engineering record will have no basis
for adjudicating between them. This is the tool-vertex trap: tooling becomes more capable while the substrate
and methodology remain inadequate, and the gap between tool capability and
substrate quality produces unreliable engineering reasoning at exactly the
workflow stages where AI participation matters most.

Table~\ref{tab:triangle} summarises the current state of each vertex and the
co-design requirement the paper argues for.

\begin{table}[ht]
\centering
\setlength{\tabcolsep}{8pt}
\renewcommand{\arraystretch}{1.4}
\small
\begin{tabular}{>{\bfseries}p{2.4cm} p{4.4cm} p{5.4cm}}
\toprule
Vertex & Current state under AI integration & Co-design requirement \\
\midrule
Language & SysML v2 textual syntax improves LLM parseability; knowledge
curation remains optional and unguided by conformance hierarchy &
Language or language extension must provide structural support for
mechanism, rationale, epistemic status, and provenance as first-class
constructs, not optional annotations \\
\addlinespace
Tool & Active investment: RAG, plug-in architectures, agentic wrappers &
Tool must enforce substrate readiness conditions specified by the
methodology, and must distinguish AI-assisted from human-authored
contributions at the artefact level \\
\addlinespace
Methodology & Unchanged: designed for human-to-human knowledge transfer;
no concept of AI as a workflow participant &
Methodology must specify substrate state requirements at each AI
engagement point, govern AI contribution validity, and provide
explicit rules for human verification and commitment \\
\bottomrule
\end{tabular}
\caption{The MBSE triangle under AI integration: current state and co-design
requirements.}
\label{tab:triangle}
\end{table}

\section{What the Co-Design Gap Looks Like in Practice}
\label{sec:failure}

Consider a programme at preliminary design review. The systems engineer has
built a SysML~v2 model, the team has recently deployed a Co-Pilot-style tool,
and a safety analyst uses it to ask a straightforward question: is the braking
requirement consistent with the assumed road condition, and where did the
stopping distance value come from?

This is exactly the kind of query the tool was purchased to answer. It is a
management-mode engagement: retrieve relevant model content, reason over it,
report back. The analyst is not asking the tool to create anything or to
understand the system as a whole. They want a specific piece of engineering
knowledge that should, in principle, already exist in the model.

The model the tool has access to is the one the systems engineer built. It is
conformant, structured, and representative of current competent MBSE practice.
Drawn from the SysML~v2 specification's own illustrative examples~\citep{sysmlv2spec},
it looks like this:

\begin{lstlisting}[style=sysml]
package Vehicle {
    private import SysML::*;
    private import ISQ::*;
    private import SI::*;
    part def Vehicle;
    concern def SafetyConcern;
    requirement def BrakingRequirement {
        subject vehicle : Vehicle;
        attribute speedLimit : ISQ::SpeedValue;
        attribute maxBrakingDistance :> ISQ::distance;
        assume constraint {
            doc /* The road condition is poor */
        }
        frame concern brakingConcern : SafetyConcern {
            :>> BrakingRequirement::maxBrakingDistance =
                    BrakingRequirement::maxBrakingDistance;
        }
    }
    part vehicle : Vehicle;
    satisfy requirement braking : BrakingRequirement by vehicle {
        :>> speedLimit = 100[km/h];
        :>> maxBrakingDistance = 10[m];
    }
}
\end{lstlisting}

\noindent
The tool reads the model. The \texttt{satisfy} relationship tells it the vehicle
shall stop from 100~km/h within 10~m. The \texttt{assume constraint} tells it the
road is poor. The \texttt{frame concern} tells it this relates to a safety concern.
There is no derivation for the 10~m value, no reference standard, no record of
who proposed it or what analysis produced it. The poor-road assumption is a
free-text doc comment with no formal link to the value it conditions. The frame
concern binding is self-referential and carries no semantic content. The following
three probes, conducted with a frontier LLM in standard configuration without
retrieval restriction,\footnote{Probes conducted with Claude (Anthropic) in standard
chat configuration. This setting gives the consumer maximum capability and no
constraint on drawing from training. The observation holds under
retrieval-augmented configurations for the same structural reason: tooling cannot
retrieve what was never recorded.} expose what this substrate produces across all
three engagement classes.

\begin{exchange}
\textbf{Probe 1 --- Management.} \textit{Why was 10~m chosen as the maximum
braking distance, and is the value consistent with the assumed road condition?}

\medskip
\noindent\textbf{Response (summarised):}
The model does not justify the 10~m value; it appears as a parameter binding
without derivation, reference standard, or rationale. Stopping from 100~km/h
in 10~m requires approximately 3.9~g of deceleration, exceeding tyre-road
friction limits on any surface. The \texttt{assume constraint} carries only a
doc comment with no formal predicate linking road condition to friction or
stopping distance. The \texttt{frame concern} block contains a self-referential
redefinition that is structurally present but semantically empty.
\end{exchange}

\noindent
The inconsistency detection is correct. Every element of that analysis was
supplied by the consumer from training; the substrate contributed values and
structural shape. The safety analyst has a useful answer. The programme has a
problem the exchange has made visible without resolving: the engineering record
contains none of the reasoning, and at a different programme, with a different
tool version or a different prompt, the same model produces a different answer.
The record cannot adjudicate between them because it was never the source.

\begin{exchange}
\textbf{Probe 2 --- Comprehension.} \textit{Summarise the design philosophy
and safety posture this requirement expresses.}

\medskip
\noindent\textbf{Response (summarised):}
The model reads as expressing a defensive safety posture: a safety-critical
concern, specified under adverse conditions, bounded by explicit parametric
values. The execution undermines the stated posture. The 10~m figure is
physically impossible, making the safety target aspirational rather than
engineered. The poor-road assumption has no formal link to the braking
calculation. The self-referential frame binding leaves the safety concern
semantically empty. The requirement is ``the shape of a defensive requirement
without the substance of one.'' The response further notes that the model
assigns no epistemic status to any of its claims: the 10~m value is presented
identically whether it is a verified engineered figure or a preliminary
placeholder, and the poor-road assumption carries no scope, no owner, and
no indication of whether it was formally agreed or informally noted. A safety
case built on this record would be asserting things the record cannot support.
\end{exchange}

\noindent
The characterisation of design philosophy is constructed entirely from structural
shape. A different consumer or a different prompt reads the same shape differently,
and the substrate has no content that adjudicates between interpretations. The
epistemically opaque substrate cannot tell any consumer which of its claims are
design targets, safety assumptions, or verified figures, so comprehension
collapses to pattern-matching on structure.

Now consider what the analyst needs to do next. The 10~m value is wrong and
someone has to decide what it should be. The safety engineer wants to propagate
the poor-road assumption through to a friction coefficient and derive a
physically achievable stopping distance. The systems engineer needs to create a
derived requirement on the brake system. Both are asking the tool to participate
in model creation, the most demanding of the three engagement classes. For that
participation to be grounded, the tool needs to know what the parent requirement's
satisfaction mechanism is, what analysis produced the current value, and what
constraints the safety case places on the derivation. None of that is in the model.

\begin{exchange}
\textbf{Probe 3 --- Creation.} \textit{Generate a derived requirement on the
brake system that contributes to satisfying this top-level requirement.}

\medskip
\noindent\textbf{Response (summarised):}
The response generates a \texttt{BrakeSystem} part, a derived
\texttt{BrakeSystemDecelerationRequirement} specialising the parent, and a
kinematic constraint relating initial speed, stopping distance, and minimum
deceleration. The minimum deceleration is computed at 38.6~m/s$^2$ from the
parent's bound values. The response explicitly flags that this is not physically
achievable, that no brake system requirement can rescue the parent's values, and
that a rigorous derivation would propagate the poor-road assumption through a
friction coefficient attribute, exposing the contradiction.
\end{exchange}

\noindent
The decomposition pattern, the kinematic constraint, the deceleration calculation,
and the identification of the contradiction are all supplied by the consumer. The
artefact provided a parent requirement and a satisfaction clause. A different
consumer generates a different derived requirement against a different satisfaction
mechanism, and the artefact has no basis for declaring one contribution coherent
and another not.

The systems engineer who built this model did not make a mistake. They followed
current practice. The methodology they worked within has no requirement to
capture rationale, derivation, or epistemic status alongside a requirement
definition. It was never designed to. The MBSE methodology was designed for
human-to-human knowledge transfer: a reviewer reading the model brings the
engineering context themselves, asks the systems engineer directly when something
is unclear, and applies professional judgement to fill the gaps. When the
consumer is a tool rather than a colleague, none of those compensation mechanisms
exist. The gap that was invisible in human-to-human practice becomes
consequential at every point where the tool engages with the model.

The substrate the tool needs is not a richer SysML model. It is a record of
the engineering decisions the process already produced but the methodology never
required to be associated with the model element:

\begin{contrastbox}
\noindent\textbf{What the substrate must carry for governed AI engagement
at this workflow stage (architecture-neutral)}

\smallskip
\noindent\textit{On the 10\,m value:} Derived from regulatory standard
ECE~R13-H under dry-road test conditions; proposed at preliminary design
review; recorded as a \emph{design target}, not a verified figure; derivation
method and source document associated with the value.

\smallskip
\noindent\textit{On the poor-road assumption:} Introduced for the safety case
worst-case scenario; scope limited to the safety analysis context; not yet
formally linked to a friction coefficient; this linkage is an open item
assigned to the brake system sub-team.

\smallskip
\noindent\textit{On the satisfaction relationship:} Satisfaction is
\emph{claimed}, not verified; the mechanism by which a brake system achieves
this stopping distance is not yet defined; no derived requirement on the
brake system exists in the model.

\smallskip
\noindent Nothing in this record specifies a SysML construct, an ontology
field, or a schema. It specifies what the substrate must make reachable, and
what epistemic status each claim must carry, for AI engagement at this
workflow stage to be grounded in the engineering record rather than
constructed from training. A methodology that requires this content to be
present before a Co-Pilot-style tool engages with this element has converted
a capability into a governed process.
\end{contrastbox}

\noindent
None of this content is invented. It existed at the moment the requirement was
written: in the review minutes, the trade study, the safety case notes, the
open-item log. The methodology never required it to be associated with the
model element. The substrate never provided a structured place for it. It
stayed in documents the tool cannot see, and the model became a structural
sketch that competent consumers fill in.

The natural response from within current practice is to require better population.
That response addresses what might be called the mandate failure: SysML~v2 places
rationale, assumption typing, and provenance in a Metadata Domain Library that tools
\emph{may} implement rather than \emph{shall}~\citep{sysmlv2spec}. Better methodology,
enforced by better tooling, would close some of that gap. But disciplined population
encounters a structural limit that no methodology requirement can override. The
\texttt{satisfy} relationship asserts satisfaction; it has no slot for the mechanism
of satisfaction, the conditions under which it holds, or the derivation chain that
justifies it. No annotation repairs this because the relationship type carries no
mechanism slot. The language records that a dependency exists, not why or how. The
co-design must therefore address the language vertex as well: either through language
extension that makes mechanism a first-class construct, or through an adjacent
substrate architecture that carries what the modelling language cannot.

This is the co-design gap in concrete form. The substrate fails because the
methodology never required it to carry what AI participation needs. The
methodology never evolved to require it because the language never provided
the constructs to express it. And the tool vertex has been built on top of
both, capable of engaging with the model in increasingly sophisticated ways,
with no change to the foundation it operates over.

\section{Three Co-Design Principles}
\label{sec:principles}

The workflow scenario in Section~\ref{sec:failure} makes the co-design gap
concrete, but concrete cases do not by themselves characterise what any solution
must achieve. We articulate three principles at the level of co-design: claims
about what the substrate-methodology pair must jointly achieve, regardless of
which architectural solution is chosen. These are not substrate properties with
methodology implications appended. They are claims about the relationship between
substrate state and methodology governance that must hold simultaneously for AI
participation to be grounded rather than fabricated.

The three principles are structured as follows. Principles~1
and~2 address the AI-as-consumer role: what the substrate must carry before AI
engages, and how the reliability of that content must be represented. Principle~2
is a necessary and specific instance of Principle~1 rather than a logically
independent claim: if the substrate carries the full context required for grounded
reasoning, that context includes the epistemic status of every claim. They are
stated separately because they identify distinct failure modes in practice and
require distinct responses from both substrate and methodology. Principle~3
addresses the AI-as-participant role: what must be true of the substrate after AI
engages, to ensure AI contributions are governed and do not degrade the record
that future engagements depend on. Together the three principles cover the full
cycle: preparation for AI engagement, reasoning over the substrate, and
contribution back to it.

These principles are independent of how AI consumption is implemented:
better retrieval, longer context windows, and more sophisticated agentic
frameworks extract more from a given substrate, but tooling cannot supply
what was never recorded, and methodology cannot govern what it never required
to be present.

\bigskip
\noindent\textbf{Co-design Principle 1 --- Engagement readiness.}
\textit{At every workflow stage where AI participates, the substrate must carry
the context required for grounded reasoning at that stage, and the methodology
must specify and enforce that substrate state as a governed condition of AI
engagement. This applies in both directions of participation: AI reading the
substrate to reason over it, and AI contributing to the substrate as part of
the workflow. Neither requirement is satisfiable independently: the substrate
cannot be designed for readiness without knowing what the workflow demands at
each stage, and the methodology cannot specify stage-level readiness conditions
without knowing what the substrate can structurally carry.}

\smallskip
This principle is what it means for AI engagement to be governed rather than
merely enabled. The distinction between element-level substrate quality and
stage-level methodology governance is important here: substrate quality is
a property of individual model elements, assessed against what each carries;
methodology governance is a property of workflow stages, specifying what
substrate state must hold programme-wide before a given class of AI engagement
is permitted to proceed. Both are necessary. A substrate whose elements are
individually complete but whose completeness is ungoverned by the methodology
will be inconsistent across programmes and across lifecycle stages. The braking
artefact in Section~\ref{sec:failure} fails this principle at both levels
simultaneously: the substrate does not carry the derivation, assumption linkage,
or satisfaction mechanism the design review workflow requires, and the methodology
specifies nothing about their presence as a condition of AI engagement at that
stage.

\bigskip
\noindent\textbf{Co-design Principle 2 --- Epistemic accountability.}
\textit{Every claim in the substrate must carry legible epistemic status, and
the methodology must govern the lifecycle of that status, specifying how claims
transition from estimated to confirmed, from conditional to universal, from
open to closed, as the programme progresses through its lifecycle stages.
Neither requirement is satisfiable independently: the substrate cannot represent
epistemic status meaningfully without a methodology that defines the status
vocabulary and the transition rules, and the methodology cannot govern epistemic
transitions without a substrate that can structurally carry them.}

\smallskip
This principle is what it means for the substrate to be honest about what it
knows at any given lifecycle stage. Confirmed measurements must be distinguishable
from estimates. Conditional claims must be distinguishable from universal ones.
A requirement value recorded as a design target at preliminary design review
and as a verified figure at critical design review carries different epistemic
weight; an AI agent reasoning over the same element at different stages must
be able to determine which it is reading. The braking artefact fails this
principle at both levels: the poor-road assumption carries no typed scope or
epistemic status, the 10~m value is presented identically whether it is a
target placeholder or an engineered figure, and the methodology specifies no
progression rules that would allow the substrate to represent the difference.
Epistemic accountability is a specific and necessary instance of engagement
readiness: a substrate that is contextually complete but epistemically opaque
is one from which an AI agent can retrieve content but cannot reason correctly
about the reliability of what it retrieves.

\bigskip
\noindent\textbf{Co-design Principle 3 --- Contribution governance.}
\textit{Every AI-assisted contribution to the substrate must be distinguishable
from human-authored content, must carry its AI-proposed epistemic status
explicitly, and must be subject to a methodology-specified review and commitment
process before it becomes part of the authoritative engineering record. Neither
requirement is satisfiable independently: the substrate cannot carry the
distinction between proposed and committed content without structural support
for contribution provenance, and the methodology cannot govern AI contributions
without a substrate that records their origin and status.}

\smallskip
This principle is what it means for AI participation to be accountable rather
than merely productive. It addresses a failure mode that Principles~1 and~2 do
not: the degradation of the substrate over time as AI-proposed content
accumulates without governed review. If the methodology does not distinguish
between an AI-proposed element and a human-verified one, and if the substrate
carries no record of that distinction, then a subsequent AI engagement over the
same substrate operates over content whose reliability is unknown. The recursive
implication is significant: ungoverned AI contributions degrade exactly the
substrate properties that Principles~1 and~2 require, compounding with each
workflow cycle. Principle~3 is therefore the completion of the co-design
framework: Principle~1 governs what the substrate must carry before AI engages,
Principle~2 governs how the reliability of that content is represented,
and Principle~3 governs what happens to the substrate after AI engages.
Together they define what it means to treat AI as a governed participant
rather than an ungoverned capability.

\bigskip
\noindent
The three probes in Section~\ref{sec:failure} demonstrate each failure mode in
turn. The absence of derivation and lifecycle provenance for the 10\,m value
fails Principle~1. The absence of typed epistemic status on the poor-road
assumption and the value binding fails Principle~2. The absence of any governed
substrate state into which a creation-mode contribution can be committed fails
Principle~3. A co-design that satisfies all three produces a substrate the next
AI engagement can trust, governed by a methodology that ensures it remains
trustworthy across the programme lifecycle.

\section{The Co-Design Problem and Its Owner}
\label{sec:codesign}

The natural owner of this co-design problem is the systems engineer. The
systems engineer's mandate, as defined in INCOSE practice, is to maintain
coherence across disciplines, system levels, and the engineering
lifecycle~\citep{walden2023incose}. No other engineering discipline holds both
the cross-cutting scope and the lifecycle mandate that substrate curation and
workflow governance jointly require. Domain specialists curate the knowledge of
their own discipline. Only the systems engineer has the mandate to curate the
substrate that connects all of them, and to specify the workflow governance that
determines when and how each discipline's AI-assisted outputs are grounded in
that substrate.

The mandate, however, is not the same as the capability or the current practice.
Three things must change concurrently.

\textbf{First, the substrate curation competency.} Producing a substrate that
satisfies all three co-design principles requires competencies not currently
standard in systems engineering education: formal knowledge representation,
schema design, provenance modelling, and disciplined epistemic classification.
These are not research specialisms to be delegated to adjacent disciplines.
They are engineering competencies to be acquired by the systems engineer
through collaboration with those disciplines, and embedded in professional
training. The call to action in
this paper therefore includes a corresponding call on the discipline to acquire
competencies it has not historically required.

\textbf{Second, the methodology revision.} Current MBSE methodologies, whether
Harmony, OOSEM, or the lifecycle processes described in the INCOSE
Handbook~\citep{walden2023incose}, are designed for human-to-human knowledge
transfer mediated by a modelling tool. They have no concept of AI as a workflow
participant: no substrate readiness conditions, no AI contribution governance,
no epistemic status transition specifications across lifecycle stages. Revising
methodology to incorporate these is not a minor amendment. It is a genuine
methodological research question, one that requires piloting, empirical
evaluation, and community debate. The research agenda proposed in
Section~\ref{sec:action} names this explicitly.

\textbf{Third, the language extension question.} The SysML~v2 extension
mechanism, particularly \texttt{metadata def} and the library conformance
structure, provides a pathway for the community to develop standard schemas
for rationale capture, assumption typing, and provenance. This is not a call
for a new language. It is a call for the OMG and the INCOSE modelling community
to treat AI-consumable substrate design as a first-class concern for language
extension work, developing standard schemas with the same rigour currently
applied to structural interoperability. The textual syntax of SysML~v2 is a
genuine enabler here: LLMs can parse, generate, and validate textual SysML
at a level of syntactic competence that diagram-based v1 did not permit.
The syntactic capability is available; the semantic schema to exploit it is not.

The reframing also changes the systems engineer's standing in an AI-augmented
enterprise. MBSE has been frequently positioned as overhead, a compliance
exercise that produces artefacts few people use~\citep{henderson2021value}.
In an enterprise where every discipline's AI-assisted output is bounded above
by the quality of the substrate it operates over, and where every AI
participant's reliability depends on what the methodology specifies about
substrate readiness, the systems engineer who owns the co-design of both holds
a position of compounding leverage over the quality of every downstream
engineering activity. The role does not need to be defended. It needs to be
redirected and equipped.

\section{A Note on Safety-Critical Systems Engineering}
\label{sec:safety}

The proposal raises a natural concern in regulated industries, where LLM use
is itself considered a risk. The concern rests on a category confusion. The
proposal concerns the engineering knowledge substrate and the systems engineer's
practice. It does not concern the certified artefacts themselves or the
certification process. The LLM is an authoring and query assistant; every
assertion that enters the substrate is made by a named human at a specific
lifecycle stage, which is precisely what Principle~2's epistemic accountability
requirement encodes, and every AI-assisted contribution is subject to the review
and commitment process Principle~3 requires. The accountability structure of
existing safety frameworks is undisturbed. Safety-critical industries already
accept that engineers use tools whose internal workings are not certified,
provided the outputs are verified by qualified humans within an approved process.
LLMs used as authoring and query assistants, governed by a methodology that
specifies substrate readiness conditions and human verification requirements,
sit squarely within this established category.

The stronger argument is that a well-governed substrate improves the
certification evidence base rather than threatening it. The most persistent
vulnerability in safety-critical systems engineering is the retrospective
documentation problem: design rationale and intent are reconstructed for
certification review rather than captured at the moment of decision. A
methodology that requires contemporaneous substrate population, with typed
epistemic status and provenance at each lifecycle stage, produces richer,
earlier, and more auditable certification evidence. The proposal does not
threaten safety-critical practice. It addresses one of its most persistent
structural weaknesses.

\section{A Call to Act}
\label{sec:action}

The three co-design principles articulated here are framing claims, not
specifications: different communities and architectures will satisfy them
differently. The community must develop, debate, and empirically test the
approaches that do so. Three specific actions are required, each directed
at a different actor.

\textbf{A substrate quality framework and reference benchmark.} The community
needs a methodology for assessing whether a given engineering artefact satisfies
all three co-design principles, independently of the architecture used to produce
it. Without a way to measure substrate quality and contribution governance, there
is no basis for evaluating progress, holding tools to account, or making substrate
readiness a legible property in programme reviews. That quality framework requires
a corresponding set of reference artefacts with known ground-truth engineering
content, against which retrieval and reasoning quality can be systematically
assessed. No such benchmark currently exists for MBSE. Establishing one would
transform the empirical evaluation of AI-assisted engineering from anecdote to
measurement, and would provide a shared basis for comparing substrate architectures
and their methodological governance. INCOSE and the wider research community are
the right bodies to lead both; the three principles proposed here are the starting
point for their operationalisation.

\textbf{A standards conversation.} The OMG and INCOSE standards processes
currently treat structural interoperability as the primary concern of modelling
language conformance. Knowledge curation is optional, unguided, and unsupported
by any standard schema. The community is called to change this: not necessarily
to mandate a specific schema, but to establish that substrate quality for AI
consumption, including the rationale, assumption, provenance, and contribution
governance constructs all three principles require, is a first-class concern
deserving standards-level attention. The SysML~v2 extension mechanism provides
a viable pathway. The standards conversation must begin before the architectural
decisions about AI integration are made by default.

\textbf{A methodology research agenda.} This is the action the community has
not yet named. Revising current methodologies to govern AI participation requires
empirical research, and the following three questions define the core of that
agenda.
First: what substrate state conditions are necessary at each AI engagement
point across the three consumption classes, and how do those conditions vary
with programme lifecycle stage? Second: what governance rules reliably
distinguish AI-assisted from human-authored contributions in a way that
survives tool change and version change in the modelling environment? Third:
how must methodology review processes evolve to include substrate readiness
assessment alongside model completeness checking, without adding overhead
that defeats adoption? This agenda cannot wait for AI tooling to mature.
The methodology must be designed to govern the participation that tooling
enables, not retrofitted to tooling that was built without it.

\bigskip
\noindent
The substrate that AI agents will consume and contribute to, on behalf of every
other discipline, is the substrate the systems engineer has always been
responsible for. The methodology that will govern those agents' participation
is the methodology the systems engineer has always been responsible for writing.
What changes is not the responsibility but the standard to which both must be met,
and the recognition that they must be met together.

\section*{Acknowledgement of AI Usage}
\addcontentsline{toc}{section}{Note on the Use of AI in Preparing This Work}

Claude (Anthropic), using both Sonnet~4.6 and Opus~4.7, was used as a
thinking partner, critical interlocutor, and editing assistant in preparing
this paper. The central argument, the identification of the co-design gap,
the three principles, the choice of worked example, and all substantive
positions are the author's own. The models contributed to stress-testing
the argument, sharpening prose, and structural editing. The three probe
responses in Section~\ref{sec:failure} were produced in a clean session
and verified against full transcripts, which are available from the author
on request. All model outputs were reviewed and accepted by the author
before entering the paper.

\bibliographystyle{plainnat}
\bibliography{refs}

\end{document}